\documentclass[12pt,preprint]{emulateapj}

\usepackage{natbib}

\slugcomment{Accepted for Publication; To appear in the ApJS Special
Issue on Spitzer, September 2004}

\shorttitle{FIR source counts with {\it Spitzer}}
\shortauthors{H. Dole et al.}

\begin{document}


\title{Far Infrared Source Counts at 70 and 160$\mu$m in {\it Spitzer} Deep Surveys}


\author{H. \ Dole\altaffilmark{1,2},
E. \ Le Floc'h\altaffilmark{1},
P. G. \ P\'erez-Gonz\'alez\altaffilmark{1}, 
C. \ Papovich\altaffilmark{1},
E. \ Egami\altaffilmark{1},
G. \ Lagache\altaffilmark{2},
A. \ Alonso-Herrero\altaffilmark{1},
C. W.\ Engelbracht\altaffilmark{1},
K. D.\ Gordon\altaffilmark{1},
D. C.\ Hines\altaffilmark{1,3},
O.\ Krause\altaffilmark{1},
K. A.\ Misselt\altaffilmark{1},
J. E.\ Morrison\altaffilmark{1},
G. H.\ Rieke\altaffilmark{1},
M. J.\ Rieke\altaffilmark{1},
J. R.\ Rigby\altaffilmark{1},
E. T.\ Young\altaffilmark{1},
L. \ Bai\altaffilmark{1},
M. \ Blaylock\altaffilmark{1},
G. \ Neugebauer\altaffilmark{1},
C. A. \ Beichman\altaffilmark{4},
D. T. \ Frayer\altaffilmark{5},
J. R. \ Mould\altaffilmark{6},
P. L. \ Richards\altaffilmark{7}
}
\altaffiltext{1} {Steward Observatory, University of Arizona, 933 N
Cherry Ave, Tucson, AZ 85721, USA}
\altaffiltext{2} {Institut d'Astrophysique Spatiale, b\^at 121,
Universit\'e Paris-Sud, F-91405 Orsay Cedex}
\altaffiltext{3} {Space Science Institute
4750 Walnut Street, Suite 205, Boulder, Colorado 80301}
\altaffiltext{4} {Michelson Science Center, CalTech, Pasadena, CA, USA}
\altaffiltext{5} {Spitzer Science Center, CalTech, Pasadena, CA, USA}
\altaffiltext{6} {NOAO, 950 N Cherry Ave, Tucson, AZ, 85719, USA}
\altaffiltext{7} {Dept. of Physics, 345 Birge Hall, Berkeley, CA, USA}



\begin{abstract}
We derive galaxy source counts at 70 and 160~$\mu$m using the
Multiband Imaging Photometer for {\it Spitzer} (MIPS) to map the
Chandra Deep Field South (CDFS) and other fields.
At 70~$\mu$m, our observations extend upward about 2 orders of
magnitude in flux density from a threshold of 15 mJy, and at
160~$\mu$m they extend about an order of magnitude upward from 50
mJy. The counts are consistent with previous observations on the
bright end. Significant evolution is detected at the faint end of the
counts in both bands, by factors of 2-3 over no-evolution models. 
This evolution agrees well with models that indicate most of the faint
galaxies lie at redshifts between 0.7 and 0.9. The new {\it Spitzer}
data already resolve about 23\% of the cosmic far-infrared background
at 70~$\mu$m and about 7\% at 160~$\mu$m. 
\end{abstract}




%
\section{Introduction}

The cosmic infrared background (CIB), relic emission of the formation
and evolution of the galaxies, peaks in the far-infrared (FIR) in the
60-200~$\mu$m wavelength range
\cite[]{puget96,hauser98,lagache99,gispert2000,hauser2001}. In the local
universe only about a third of the extragalactic emission is released
in the FIR \cite[]{soifer91}. However, the CIB FIR
peak accounts for more than half of the total optical/infrared
background, indicating strong evolution of galaxy properties toward
high FIR output in the past. Characterizing the galaxies responsible
for most of the CIB is therefore an important goal of cosmological
surveys. Galaxy counts (or number counts) provide a powerful tool to
investigate the evolution of the galaxies and their contribution to
the CIB. 

The cryogenic infrared space missions {\it IRAS} (Infrared Astronomical
Satellite) and {\it ISO} (Infrared Space Observatory) \cite[for
reviews]{genzel2000,dole2003a} provided important data on source
counts at 60~$\mu$m
\cite[]{hacking91} and at 
100~$\mu$m \cite[]{rowan-robinson86}
, and more recently at
90~$\mu$m \cite[]{kawara98,efstathiou2000,juvela2000,linden-voernle2000,matsuhara2000,rodighiero2003,kawara2004}
and 170~$\mu$m \cite[]{kawara98,puget99,matsuhara2000,dole2001,kawara2004}.
Mid-infrared (MIR) observation with ISOCAM at 15$\mu$m \cite[]{elbaz99} 
are also of great interest since they are believed to resolve a
significant fraction of the CIB into sources \cite[]{elbaz2002a}. The
{\it Spitzer Space Telescope} \cite[]{werner2004} provides the ability for
much deeper and wider-area surveys from 3.6 to 160~$\mu$m.
This paper investigates source counts at 70 and 160~$\mu$m from {\it
Spitzer}.  
A companion paper addresses the MIR source counts at 24~$\mu$m
\cite[]{papovich2004}. The three-band source counts are the basis of
new phenomenological models by \cite{lagache2004}.

%
\section{Observations and Data Reduction}
\label{sect:obs}

Observations were carried out with the Multiband Imaging Photometer
for {\it Spitzer} \cite[]{rieke2004} in the {\it Chandra} Deep Field
South (CDFS) and the Bo\"otes field corresponding to the NOAO Deep
Wide Field Survey NDWFS \cite[]{jannuzi99}; We also used an
engineering MIPS observation of the Marano field. The observational
mode (scan map) provides multiple sightings of each source, typically
10 and 60 at 70$\mu$m in the Bo\"otes and CDFS, respectively. However,
at 160~$\mu$m, the number of sightings is only typically 2 in Bo\"otes
and 12 in CDFS. See Table~\ref{fields} and
\cite{papovich2004} for details.

\begin{figure}
\plotone{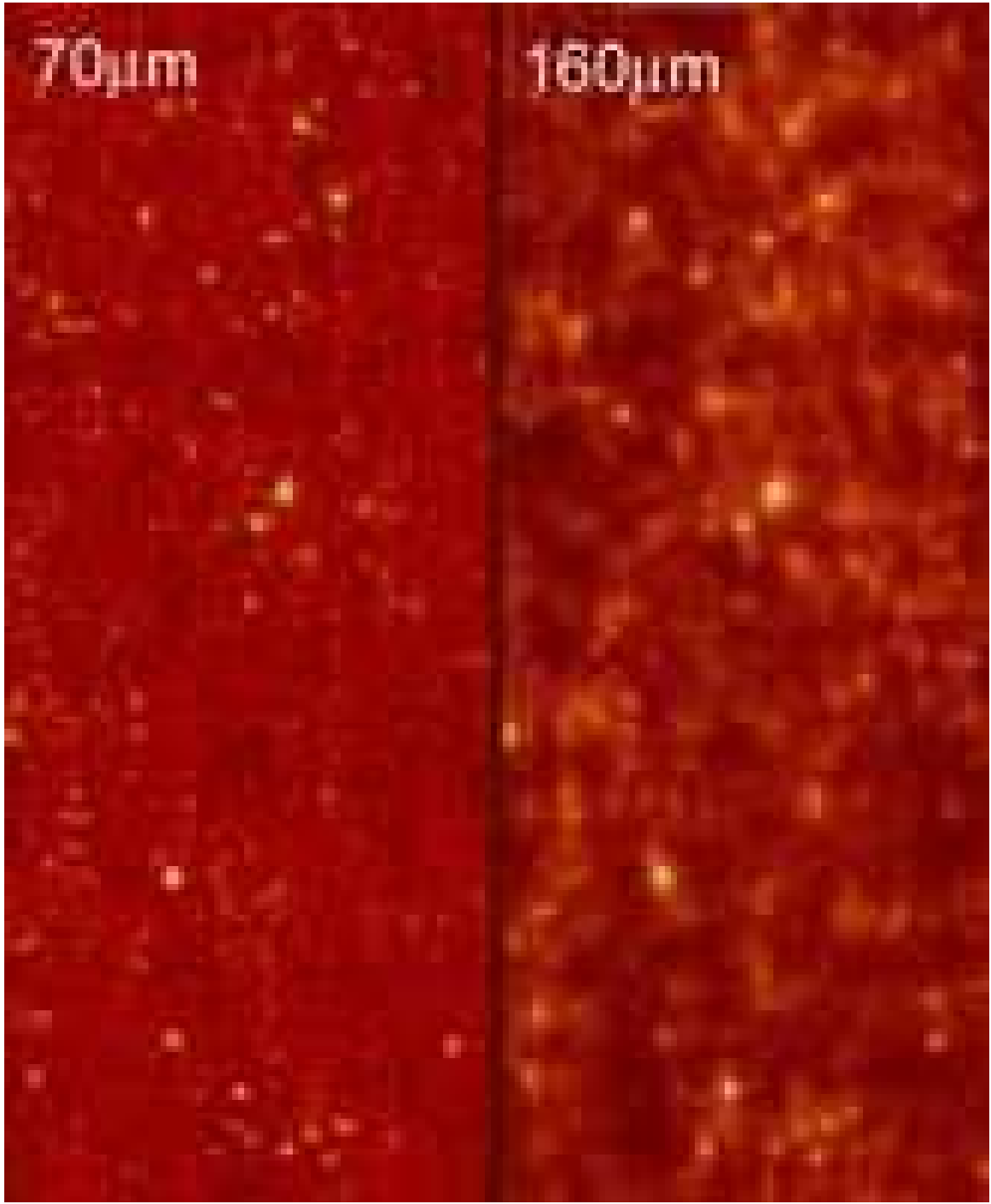}
\caption{MIPS observations of the Chandra Deep Field South at 70 and
160~$\mu$m. The field covers an area of $25' \times 1^{\circ}$.
\label{fig:f1}}
\end{figure}

The data were reduced with the Data Analysis Tool 
\cite[]{gordon2004}, from the raw data (ramps) to the final coadded
mosaics. The illumination corrections were derived from the data
themselves. At 70~$\mu$m, the data have been median-filtered in the
time domain before mosaicking. Note that data from Ge:Ga detectors are
always challenging to process; but with MIPS, most of the difficulties
are overcome with frequent calibrations (stimulator flashes), that
track the responsivity variations. Nevertheless, the noise properties
at faint fluxes  are still being investigated at both 70 and 160~$\mu
m$. In this work, we will adopt conservative detection limits.
A future paper will address extracting the ultimate sensitivity from
these data.
Sample images in the CDFS are shown in Figure~\ref{fig:f1}.

%
\begin{deluxetable}{cccccccc}
\tablewidth{0pt}
\tablenum{1}
\tablecaption{Log of Observations}
\tablehead{
\colhead{Field}                              &
\colhead{MIPS}                               &
\multicolumn{3}{c}{70 $\mu$m}                &
\multicolumn{3}{c}{160 $\mu$m}               \\
\colhead{Name}                               &
\colhead{AOT\tablenotemark{b}}               & 
\colhead{Area}                               &     
\colhead{t$_{int}$\tablenotemark{c}}         &  	
\colhead{S$_{cut}$\tablenotemark{d}}         &
\colhead{Area}                               &     
\colhead{t$_{int}$\tablenotemark{c}}         &
\colhead{S$_{cut}$\tablenotemark{d}}         \\
\colhead{\tablenotemark{a}}                  &
\colhead{}                                   &
\colhead{$[deg^2]$}                          &  	
\colhead{[s]}                                &
\colhead{[mJy]}                              & 
\colhead{$[deg^2]$}                          &  	
\colhead{[s]}                                &
\colhead{[mJy]}                                                                              
}
\startdata
Bo\"otes & med. & 8.75 & 40  & 80 & 7.70 & 8   & \nodata\tablenotemark{e}\\
Marano & slow & 0.42 & 100 & 25 & 0.31 & 20  & 50 \\
CDFS   & slow & 0.67 & 600 & 15 & 0.54 & 120 & 50 \\
\enddata
\label{fields}
\tablenotetext{a}{ see Papovich et al.\ (2004) for details on fields}
\tablenotetext{b}{ scan map mode}
\tablenotetext{c}{ per sky pixel}
\tablenotetext{d}{ flux density at which catalog was cut}
\tablenotetext{e}{ field data not used at 160~$\mu$m}
\end{deluxetable}

%
\section{Photometry \& Catalogs}
\label{sect:photometry}

To control the sample and the selection function, we accepted source
detections only where the redundancy was high 
(typically 80\% or more of the mean weight), avoiding the edges
and the noisiest areas of the images. The resulting positions were fed to
DAOPHOT \cite[]{stetson87} in IRAF\footnote{IRAF is distributed by the
National Optical Astronomy Observatories, which are operated by the
Association of Universities for Research in Astronomy, Inc., under the
cooperative agreement with the National Science Foundation}
for PSF fitting. We checked that the residual maps were indeed free of
sources.  

At 70~$\mu$m the photometric calibration is derived from many
observation campaigns, and its uncertainty is conservatively estimated
at the order of 20\%. We use only detections at 15 mJy and brighter in
the CDFS, 25 mJy and brighter in Marano, and 80 mJy and brighter in
the Bo\"otes field. These flux levels are determined using the sharp
decrease in the counts due to the incompleteness effect. At
160~$\mu$m, the calibration is based on a combination of observations
of standard stars, asteroids, and comparisons with measurements with
other FIR missions ({\it ISO}, {\it COBE}, and modeling including IRAS
measurements). It is also estimated to be accurate to about 20\%.  We
have included in our counts only objects of 50mJy and brighter in the
CDFS, and Marano (levels determined using incompleteness as at
70~$\mu$m).  Because of the low redundancy level of the 160~$\mu$m
data in the Bo\"otes field, we postpone using it for a later paper.

Catalogs were produced separately at each wavelength; surveys at
each wavelength are thus {\it unbiased}.
Detected FIR sources in the CDFS sort as follows.
For sources selected at 70~$\mu$m:
92\% have a 24~$\mu$m ID;
54\% have a 160~$\mu$m ID (same with 24 and 160).
For sources selected at 160~$\mu$m:
98\% have a 24~$\mu$m ID;
43\% have a 70~$\mu$m ID (same with 24 and 70).

%
\section{Source counts}
\label{sect:counts}

At 70~$\mu$m, 131 sources were detected down to 15 mJy
in the CDFS, 55 sources down to 25 mJy in Marano, and
117 down to 80 mJy in the Bo\"otes.
At 160~$\mu$m, down to 50 mJy, 123 sources were detected
in the CDFS, and 89 sources in Marano.
The source density corresponds to about 150 beams per source
at 70~$\mu$m and 15 at 160~$\mu$m, using the definition of
\cite{helou90}.

\begin{figure}
\begin{center}
\includegraphics[width=11cm]{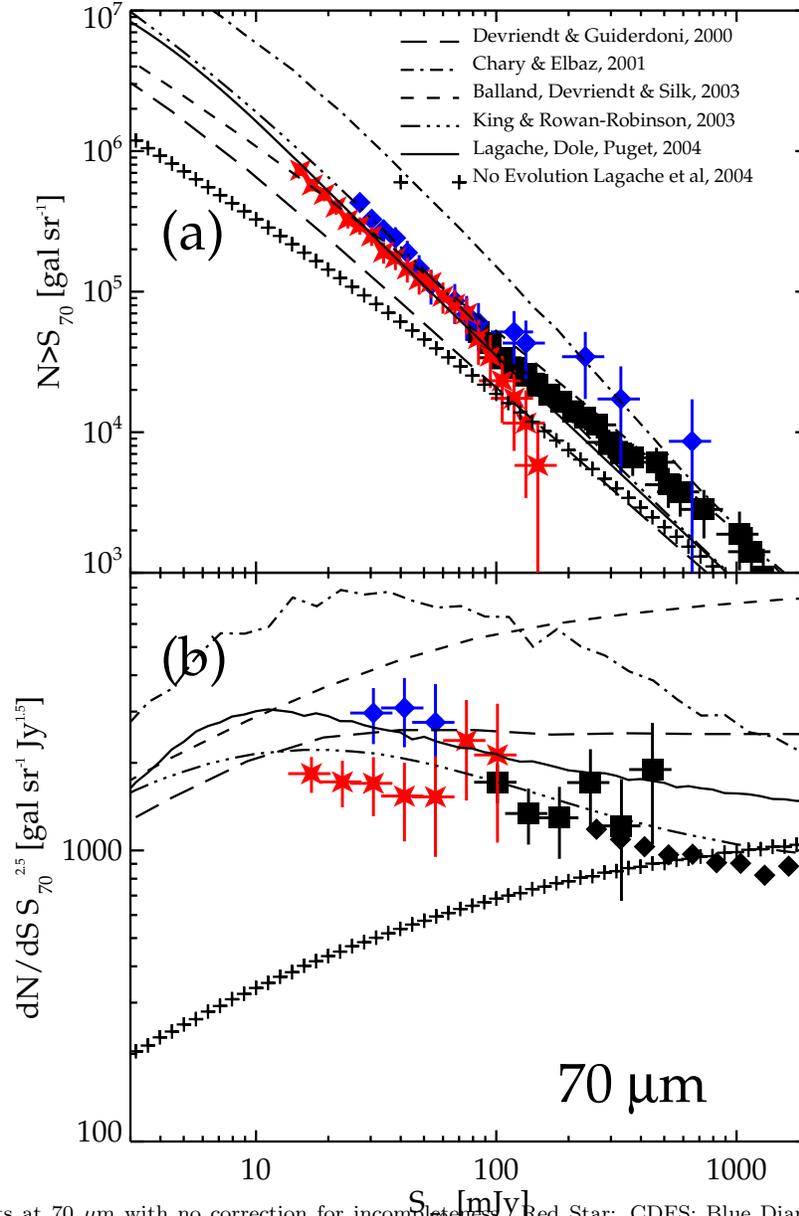}
\end{center}
\vspace{-2cm}
\caption{Source Counts at 70~$\mu$m with no
correction for incompleteness.
Red Star: CDFS;
Blue Diamond: Marano;
Black Square: Bo\"otes Field;
Black Diamond: IRAS 60~$\mu$m counts from \cite{lonsdale90} converted
at 70~$\mu$m.
Top panel (a): Integral Source Counts.
For clarity, photometric
uncertainty is only shown for $S_{70}>100$ mJy.
Bottom panel (b): Differential Source Counts.
Models are also plotted:
long dash: \cite{devriendt2000};
dash-dot: \cite{chary2001};
dash: \cite{balland2003}; 
dash-dot-dot-dot: \cite{king2003}; 
solid line: \cite{lagache2004};
plus: No Evolution Model.
\label{fig:sourcecounts_int_and_diff_legend_0070}}
\end{figure}

\begin{figure}
\begin{center}
\includegraphics[width=10cm]{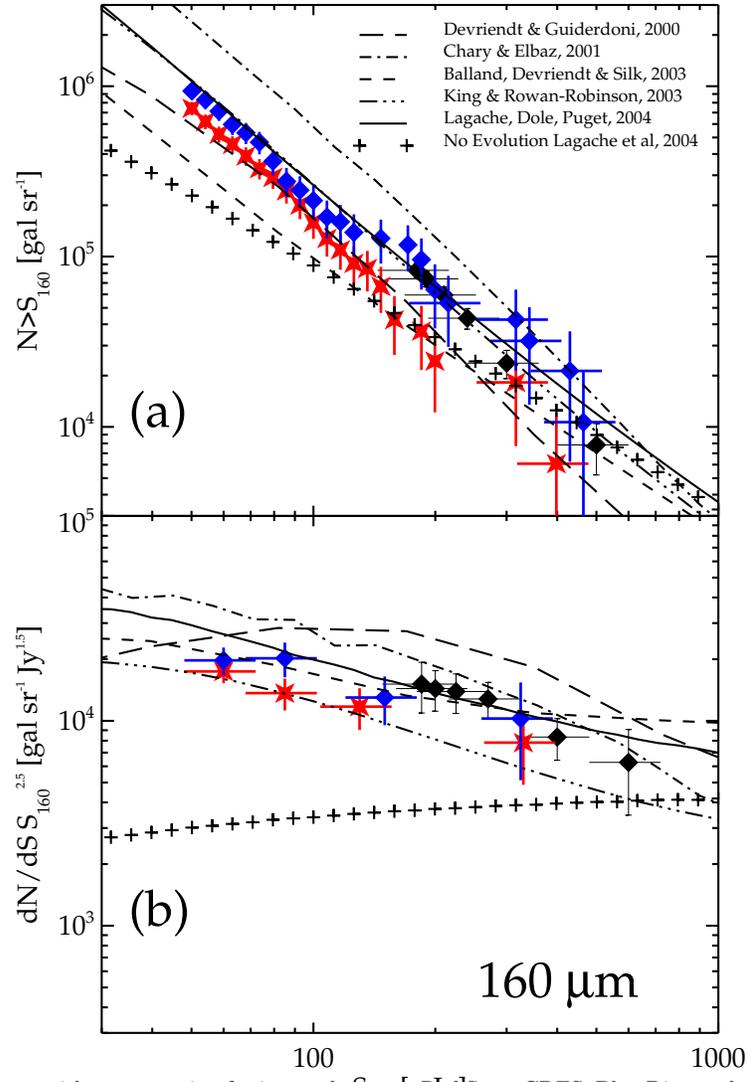}
\end{center}
\vspace{-2cm}
\caption{Source counts at 160~$\mu$m with no
correction for incompleteness.
Red Star: CDFS;
Blue Diamond: Marano field;
Black Diamond: ISO FIRBACK 170~$\mu$m counts from \cite{dole2001}.
Top panel (a): Integral Source counts. For clarity, photometric
uncertainty is only shown for $S_{160}>200$ mJy.
Lower panel (b): Differential Source counts.
Models are also plotted with the same symbols as
Figure~\ref{fig:sourcecounts_int_and_diff_legend_0070}.
\label{fig:sourcecounts_int_and_diff_legend_0160}} 
\end{figure}

Source counts are given in integral form
(Figure~\ref{fig:sourcecounts_int_and_diff_legend_0070}a and
\ref{fig:sourcecounts_int_and_diff_legend_0160}a) and
differential form, divided by the Euclidean component
(Figure~\ref{fig:sourcecounts_int_and_diff_legend_0070}b and
\ref{fig:sourcecounts_int_and_diff_legend_0160}b) at 70
and 160~$\mu$m, respectively.
Notice that we did not correct for incompleteness.
Error bars on counts are $1\sigma$ Poisson uncertainty.
Bins with less than four sources have not been displayed for clarity,
since their significance is low, with uncertainties of 50\% or higher.
Also, photometric uncertainty have been displayed at high flux only.
In order to visualize the contribution from each field, source counts
have not been merged and have been overplotted.
One should keep in mind that MIPS source counts will eventually go
deeper and will be corrected for incompleteness.

The observed fields nicely complement each other in terms of area and
depth. This allows us to probe a flux range of almost 2 orders of
magnitude at 70~$\mu$m. One order of magnitude is covered at 160
$\mu$m.
It is possible to check consistency and the cosmic variance in the
common flux density range. At 70~$\mu$m, in the range 25 to 100 mJy
where three fields overlap in flux density, the differential counts
are almost consistent within the error bars.
At 160~$\mu$m, in the range 100 to 300 mJy, the differential counts
are consistent within the error bars. At both wavelengths,
number counts in CDFS appear consistently lower than in Marano.

%
\section{Discussion}
\label{sect:discussison}

\subsection{70~$\mu$m} 

The MIPS 70~$\mu$m counts show a great consistency with the {\it IRAS}
60~$\mu$m counts of \cite{lonsdale90} converted at 70~$\mu$m using
$\nu_{60} S_{60} = \nu_{70} S_{70}$.

A selection of recent models is shown in
Figure~\ref{fig:sourcecounts_int_and_diff_legend_0070}, including a
non-evolution scenario.
The most striking result is the strong excess of MIPS 70~$\mu$m
sources compared to the non-evolution model: a factor of 3 at around
20-30 mJy. Strong evolution had been reported previously at 60 and
90 $\mu$m, and these data provide unambiguous confirmation.

\begin{figure}
\begin{center}
\includegraphics[width=15cm]{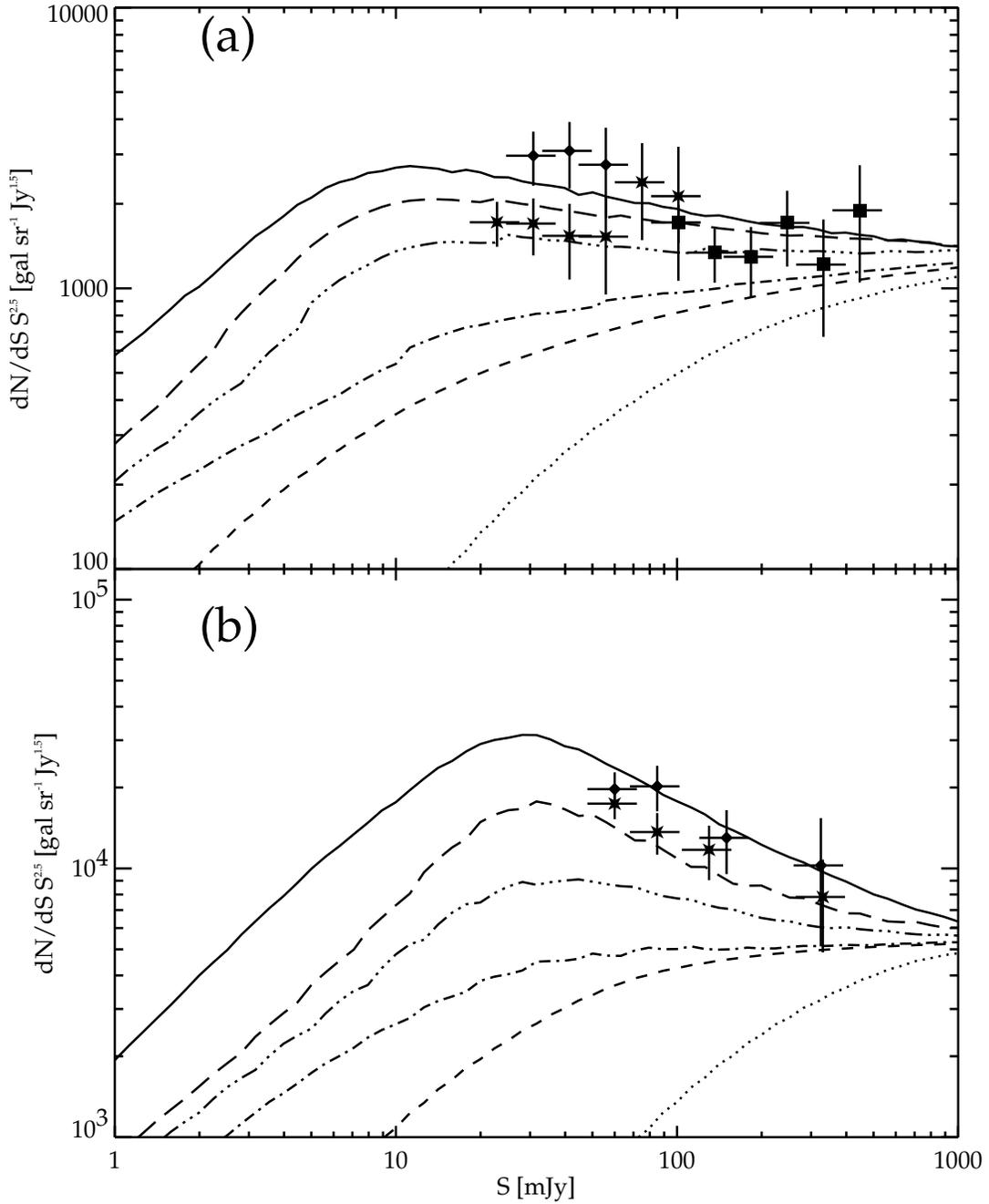}
\end{center}
\vspace{-2cm}
\caption{Cumulative contributions to the Differential Source Counts at
70~$\mu$m and 160~$\mu$m of galaxies, as a function of redshift,
form the model of \cite{lagache2004}.
Top panel (a): 70~$\mu$m.
Bottom panel (b): 160~$\mu$m.
Symbols are described in
Figures~\ref{fig:sourcecounts_int_and_diff_legend_0070}
and \ref{fig:sourcecounts_int_and_diff_legend_0160}.
Galaxies contributing to the counts at Redshifts
0.1 (dot), 0.3 (dash), 0.7 (dot-dash), 0.9 (dot-dot-dot-dash)
and 1.1 (long dash). The total contribution is the upper solid line.
\label{fig:zcontrib_counts}} 
\end{figure}

Two models lie close to the data:
\cite{king2003} and \cite{lagache2003,lagache2004}.
These models, developed to fit observables mostly from {\it IRAS},
{\it ISO} and SCUBA surveys as well as the CIB spectral energy
distribution (SED), are based on a strong evolution of luminous (and
ultra-luminous) infrared galaxies (LIRG and ULIRG). The latter model
predicts a peak in the redshift distributions of resolved sources at
70~$\mu$m near $z \sim 0.7$ \cite[]{dole2003}.
Figure~\ref{fig:zcontrib_counts}a shows the galaxy contribution to the
differential counts, as a function of redshift, from the
\cite{lagache2004} model.
Between $\sim$5 and $\sim$100 mJy, sources at $0.7 \le z \le 0.9$
contribute the most to the counts. At brighter fluxes (reached by {\it
IRAS} and {\it ISO}), contributions from local galaxies are more
important.

The source counts integrated at 70~$\mu$m correspond to a brightness
of 0.022 MJy/sr or 0.95 nW/m$^2$/sr. The value of the CIB at this
wavelength is not known accurately owing to contamination by zodiacal
light. If we use the CIB value from the model of
\cite{lagache2004}, the MIPS counts show that at 70~$\mu$m about
23\% of the 70~$\mu$m CIB is already resolved. 

\subsection{160~$\mu$m}

At 160~$\mu$m, the comparison with the {\it ISO} FIRBACK 170~$\mu$m survey
\cite[]{dole2001} shows that the counts are consistent and within the
error bars in the whole common range, 180-300 mJy. Other {\it ISO}
observations \cite[]{matsuhara2000,kawara2004} agree as well.

Figure~\ref{fig:sourcecounts_int_and_diff_legend_0160}
overplots the same models as at 70~$\mu$m. The evolution detected at
170~$\mu$m is confirmed at 160~$\mu$m, down to fainter levels. At
about 100 mJy, an excess of sources by more than a factor of 2 is
observed compared to a non-evolution scenario.
Interestingly, as for 70~$\mu$m, the observed evolution is better
fitted by the models of \cite{lagache2004} and \cite{king2003}.
The observed slope also agrees with \cite{mould2003}.

We have constrained the bright end of galaxy number counts
at 170~$\mu$m by using data from the ISOPHOT Serendipity Survey
(ISOSS). ISOSS provides a total sky coverage of 15\% and is virtually
complete at a flux density level of $S_{\rm 170\,\mu m} = 50\,{\rm Jy}$.
Based on all optically identified galaxies detected by ISOSS
\cite[]{krause2003,stickel2004} we have derived an integral number density
of $ n(S_{\rm 170\,\mu m} > 50\,{\rm Jy}) = 14 \pm 3$ gal.sr$^{-1}$
galaxies at high Galactic latitudes.  This point is perfectly matched
by the model of \cite{lagache2003,lagache2004}.
 
The most striking result of the models that fit the number counts is 
the existence of two regimes in flux density.
In the {\it ISO} range (fluxes above 200 mJy) most of the sources
contributing to the counts are local; this is confirmed by observation
\cite[]{patris2003a}. At fainter fluxes, between
$\sim$10 and 200 mJy, the counts should be dominated by a population located at
redshifts between 0.7 and 0.9.

The source counts integrated at 160~$\mu$m correspond
to a brightness of 0.07 MJy/sr or 1.4 nW/m$^2$/sr.
The CIB value at this wavelength is 1 MJy/sr \cite[]{lagache2000};
the MIPS counts show that at 160~$\mu$m about 7\% of the 
CIB is resolved. Since these counts are preliminary, are not corrected
for incompleteness, and are subject to cosmic variance, we anticipate
that the actual value might be higher. 

\subsection{Concluding Remarks}

The first MIPS far-infrared source counts, spanning
about 2 orders of magnitude in flux density at 70~$\mu$m (and one at
160~$\mu$m), are consistent on the bright end with previous
observations, and show on the faint end unambiguous evolution. Models
predict that most of the sources lie at $z \sim 0.7$ with a tail up to
$z \sim 2$ \cite[]{dole2003,lagache2003,lagache2004}.
This work and companion papers about source counts at 24~$\mu$m
\cite[]{papovich2004}, about confusion at 24, 70 and 160~$\mu$m
\cite[]{dole2004a}, and about the interpretation of these new
data from the {\it Spitzer} cosmological surveys
\cite[]{lagache2004}, bring new light on the statistical properties of
galaxies in an unexplored regime in flux density, and likely in a
critical region of redshift space (up to redshifts $z \sim 2$) in the
FIR \cite[]{egami2004,lefloch2004}.

%
\acknowledgments
This work is based on observations made with the {\it Spitzer} Space
Telescope, which is operated by the Jet Propulsion Laboratory,
California Institute of Technology under NASA contract 1407.
Support for this work was provided by NASA through contract
960785 issued by JPL/Caltech.
We warmly thank J. Cadien and J-L. Puget, and the IRS Team for
providing us with the Bo\"otes data.

%


%

%

%

%

\end{document}